\documentclass[twocolumn,preprintnumbers,prd,superscriptaddress,nofootinbib,showpacs]{revtex4}

\usepackage{amsmath}
\usepackage{amsfonts}
\usepackage{amssymb}
\usepackage{graphicx}
\usepackage{color}
\usepackage{hyperref}
\usepackage{booktabs}
\usepackage{hyperref}
\usepackage{cleveref}
\usepackage{color}
\usepackage{braket}
\usepackage{mathtools}
\usepackage[rightcaption]{sidecap}
\usepackage{subfigure}
\usepackage{tensor}

\newcommand{\be}{\begin{equation}}
\newcommand{\ee}{\end{equation}}
\newcommand{\bs}{\begin{split}}
\newcommand{\bea}{\begin{eqnarray}}
\newcommand{\eea}{\end{eqnarray}}

\newcommand{\bes}{\begin{subequations}}
\newcommand{\ees}{\end{subequations}}

\begin{document}

\title{Entanglement production in Einstein-Cartan theory}

\author{Alessio Belfiglio}
\email{alessio.belfiglio@studenti.unicam.it}
\affiliation{Scuola di Scienze e Tecnologie, Universit\`a di Camerino, Via Madonna delle Carceri 9, 62032 Camerino, Italy.}

\author{Orlando Luongo}
\email{orlando.luongo@unicam.it}
\affiliation{Scuola di Scienze e Tecnologie, Universit\`a di Camerino, Via Madonna delle Carceri 9, 62032 Camerino, Italy.}
\affiliation{Dipartimento di Matematica, Universit\`a di Pisa, Largo B. Pontecorvo 5, Pisa, 56127, Italy.}
\affiliation{NNLOT, Al-Farabi Kazakh National University, Al-Farabi av. 71, 050040 Almaty, Kazakhstan.}

\author{Stefano Mancini}
\email{stefano.mancini@unicam.it}
\affiliation{Scuola di Scienze e Tecnologie, Universit\`a di Camerino, Via Madonna delle Carceri 9, 62032 Camerino, Italy.}
\affiliation{Istituto Nazionale di Fisica Nucleare, Sezione di Perugia, Via Alessandro Pascoli 23c, 06123 Perugia, Italy.}

\begin{abstract}
We study the entanglement production for Dirac and Klein-Gordon fields in an expanding spacetime characterized by the presence of torsion. Torsion is here considered according to the Einstein-Cartan theory with a conformally flat Friedmann-Robertson-Walker spacetime. In this framework, torsion is seen as an external field, fulfilling precise constraints directly got from the cosmological constant principle. For Dirac field, we find that torsion increases the amount of entanglement. This turns out to be particularly evident for small values of particle momentum. We discuss the roles of Pauli exclusion principle in view of our results, and, in particular, we propose an interpretation of the two maxima that occur for the entanglement entropy in presence of torsion. For Klein-Gordon field, and differently from the Dirac case, the model can be exactly solved by adopting the same scale factor as in the Dirac case. Again, we show how torsion affects the amount of entanglement, providing a robust physical motivation behind the increase or decrease of entanglement entropy. A direct comparison of our findings  is also discussed in view of previous results derived in absence of torsion. To this end, we  give prominence on  how our expectations would change in terms of the coupling between torsion and the scale factor for both Dirac and Klein-Gordon fields.
\end{abstract}


\maketitle

\section{Introduction}

Our current understanding of the universe is undergoing a new revolutionary phase in which observations provide precise measurements that fix bounds on the cosmological parameters characterizing the standard cosmological model \cite{uno,due}. In this respect, the interplay between quantum world and gravitation is an ambitious challenge for theoretical physics as it sheds light on how early-phases evolve when general relativity breaks down \cite{tre}. Applications to quantum gravity could open new windows on the properties of the initial singularity, inflation \cite{pre4} and likely on the existence of both dark energy \cite{quattro} and matter \cite{cinque}. It is therefore of interest to explore different scenarios, choosing them through helpful guiding principles that make use of a
minimal number of assumptions and ingredients. All these scenarios lie on postulating the cosmological principle, i.e. essentially the key ingredient that assumes the universe to be homogeneous and isotropic \cite{sette}. In this framework, it is interesting to consider the Einstein-Cartan (EC) theory \cite{cartan0} in which the role of torsion represents the simplest modification of Einstein's gravity \cite{cartan1,cartan2}. The torsion is assumed not to vanish as in general relativity, enabling one to match its existence to particle spin. Here, spin plays a dynamical role \cite{23,24,25}, in fact we assume the  torsion field to couple with the spin of particles, giving rise to interacting terms that act on the overall dynamics. For these considerations, it is natural to work on particle production and on its applications to quantum cosmology when EC theory is accounted.

Indeed, an intriguing topic that is currently object of speculation in cosmology is represented by entanglement production in asymptotic phases \cite{cosmoentanglement1}. Entanglement is a fundamental property of quantum systems implying the existence of global states of composite systems which cannot be written as a product of the states of individual subsystems \cite{2}. It recently started to be a resource in quantum information theory, with several applications that span from quantum communication \cite{3}, quantum cryptography \cite{4}, quantum teleportation \cite{5} up to quantum computation \cite{6} and, more recently, to its characterization in relativistic frameworks \cite{7,8}, such as in curved spacetime \cite{9,10,11,12}.

Spacetime curvature  has nontrivial effects on quantum fields living on the
spacetime when compared with their flat-spacetime counterparts \cite{cosmoentanglement2}. This is especially interesting in the case of dynamical spacetime backgrounds because  the gravitational interaction may induce quantum correlations in the field state in scenarios such as expanding universes
\cite{9,10,11,12,13,14}. This is related to the long known phenomenon of particle-antiparticle production from vacuum \cite{particle}. It was shown that Dirac and KG field have different momentum distribution of entanglement \cite{13}.
Within the Dirac field no qualitative difference appear in the dependence of entanglement from the number of created particles at fixed momentum
in going from $1+1$ spacetime to $3+1$ spacetime, hence including spin \cite{18,19}. However all these studies were confined to torsion-less spacetimes.
The inclusion of torsion can shed further light on the differences between the entanglement of bosonic and fermionic fields, and in particular concerning the role the spin plays in its generation.

In this paper, we face the problem of investigating entanglement production for bosonic and fermion particles in an expanding spacetime with the presence of nonzero torsion. To do so, we consider the Dirac and Klein-Gordon (KG) field and  the role of spin in view of EC theory. In this respect, we investigate entanglement for Dirac and KG field within the EC theory. Thus, assuming the cosmological principle to hold, we adopt the Friedmann-Friedmann-Robertson-Walker (FRW) spacetime, fulfilling constraints for the torsion provided by recent observations \cite{26} and we notice  the effect of the torsion appears in the dependence of the particle density from momentum. Thus, invoking a generic torsion source, reinterpreted as an external geometrical source, we describe how Dirac field are minimally coupled to torsion and how a non-minimal coupling of torsion to KG field is plausible, both introducing significant curvature effects. Afterwards, we show how the presence of torsion affects entanglement, in both the cases. In particular, we show how to get from the Dirac equation physical solutions in presence of torsion in particular spacetime regions. These solutions are not analytical as well as the corresponding entanglement entropy. However, by assuming small corrections due to torsion we get approximate classes of solutions that resemble previous results developed in the literature where torsion was not  taken into account. As a consequence, we underline how torsion deviates the standard expectations and under which conditions  torsion can increase or decrease particle and entanglement productions. The opposite happens for KG field. There, although the torsion effect is modeled in a more complicated way, i.e. adopting two sources instead than one as for Dirac, exact solutions can be argued. According to our findings, we show under which properties torsion can increase the amount of entanglement and how much it is mode dependent. We underline the entanglement increase is particularly marked  for small values of the particle momentum. Consequences in cosmology and imprints on observations are discussed. In particular, we interpret our findings in view of the Pauli exclusion principle, explaining the presence of a relative maximum for the Dirac field.

The paper is structured as follows. In Sect. \ref{Basics}, basic notions of EC theory are reported, giving emphasis on how to fuel torsion by means of the most generic approach. Thus, in Sect. \ref{sez3} we discuss how to relate EC gravity to Dirac and how torsion modifies the entanglement production. The same is faced for KG field in Sect. \ref{sez4}. A comparison of both the frameworks is extensively discussed throughout Sect. \ref{sez5}. In the same section, we also give a physical interpretation of our results and we stress how to relate our torsion fields to Pauli exclusion principle. Finally, in Sect. \ref{sez6} we discuss conclusions and perspectives of our work.

\section{The EC theory}
\label{Basics}

The EC theory can be introduced starting from the action
\be \label{1}
\mathcal{L}_{EC}= -\frac{1}{2\kappa c} \int R(\Gamma) \sqrt{-g}\ d^4x+ \int \mathcal{L}_m \sqrt{-g}\  d^4x,
\ee
where $\kappa\equiv8\pi G$ and $g$ is the determinant of the spacetime metric tensor $g_{\mu \nu}$. The Lagrangian $\mathcal{L}_m$ represents a generic matter contribution. This action is defined in a spacetime with curvature and torsion, usually called Riemann-Cartan (RC) spacetime. The curvature scalar $R(\Gamma):=g^{\mu \nu}R_{\mu \nu}$ is constructed out of the Ricci-Cartan tensor $R_{\mu \nu}(\Gamma) \equiv R^{\alpha}_{\mu \alpha \nu}(\Gamma)$, while the torsion tensor $T^{\alpha}_{\hphantom{\alpha}\mu \nu}$ is defined as the antisymmetric part of the affine connection
\be \label{2}
T^{\alpha}_{\hphantom{\alpha}\mu \nu}:= \Gamma^{\alpha}_{\hphantom{\alpha}[\mu \nu]}= \frac{1}{2}\left( \Gamma^{\alpha}_{\hphantom{\alpha} \mu \nu}- \Gamma^{\alpha}_{\hphantom{\alpha} \nu \mu} \right).
\ee
Accordingly, the affine connection can be written as the sum of two contributions \cite{27}:
\be \label{3}
\Gamma^{\alpha}_{\hphantom{\alpha} \mu \nu}= \tilde{\Gamma}^{\alpha}_{\hphantom{\alpha} \mu \nu}+ K^{\alpha}_{\hphantom{\alpha} \mu \nu},
\ee
where $\tilde{\Gamma}^{\alpha}_{\hphantom{\alpha} \mu \nu}$ is the usual Levi-Civita spin connection of general relativity and $K^{\alpha}_{\hphantom{\alpha} \mu \nu}$ is the contorsion tensor, related to torsion via the formula
\be \label{4}
K_{\alpha \mu \nu}:= T_{\alpha \mu \nu}+ 2T_{(\mu \nu) \alpha}.
\ee
In the EC theory we deal with a set of two field equations: the \textit{first Einstein-Cartan equation} relates the curvature of spacetime to the energy-momentum density of matter, described by the tensor $T_{\mu \nu}$. This equation maintains the same form of standard general relativity, i.e. $
R_{\mu \nu}-\frac{1}{2}Rg_{\mu \nu}= \kappa T_{\mu \nu}$, but without having the \emph{a priori} symmetry of both the Ricci-Cartan and energy-momentum tensors.

The \textit{second Einstein-Cartan equation} couples the spacetime torsion to the matter spin. It can be written as
\be \label{6}
T^{\alpha}_{\hphantom{\alpha} \mu \nu}-T_{\mu}\delta^{\alpha}_{\nu}+ T_{\nu} \delta^{\alpha}_{\mu}=- \frac{\kappa}{2}s_{\mu \nu}^{\hphantom{\mu} \hphantom{\nu} \alpha},
\ee
where
\be \label{7}
s_{\alpha}^{\hphantom{\alpha} \mu \nu}= \frac{2}{\sqrt{-g}} \frac{\delta \mathcal{L}_m}{\delta K^{\alpha}_{\hphantom{\alpha}\mu \nu}}
\ee
is the spin tensor of matter.

\subsection{The cosmological principle in theories with torsion}
Now we want to specify to the case of a spatially homogeneous and isotropic spacetime, described by the conformal FRW line element
\be \label{8}
ds^2= a^2(\tau)(-d\tau^2+dx^2+dy^2+dz^2).
\ee
Here $a(\tau)$ is the scale factor, determining the spacetime expansion rate, while $\tau$ is the conformal time, related to the cosmological time t by $\tau= \int a^{-1}(t) dt$.
Given the high symmetry of such a spacetime, the torsion tensor has to satisfy certain constraints. We follow the ansatz of \cite{26} and assume that the only non-zero components of the torsion tensor are\footnote{We are assuming that torsion is invariant under conformal transformations. For an introduction to conformal properties of torsion see, for example, \cite{29}.}
\be \label{9}
T_{\alpha \mu \nu}=f(\tau) \epsilon_{\alpha \mu \nu},\ \ \ \ \ T^{\alpha}_{\hphantom{\alpha} \mu 0}=h(\tau)\delta^{\alpha}_{\mu},
\ee
where $f(\tau)$ and $h(\tau)$ are arbitrary functions of the conformal time, while $\epsilon_{\alpha \mu \nu}$ and $\delta^{\alpha}_{\mu}$ are the three-dimensional Levi-Civita and Kronecker symbols, respectively. Using the definition \eqref{4}, from \eqref{9} we obtain
\be \label{10}
K_{\alpha \mu \nu}=f(\tau) \epsilon_{\alpha \mu \nu},\ \ \ \ \ K_{0 \mu \nu}=-K_{\mu 0 \nu}= 2 h(\tau) g_{\mu \nu}.
\ee
This ansatz is valid for any gravity theory in a RC spacetime, if one applies the cosmological principle to the torsion tensor. In doing this, we drop any assumptions about the source of torsion.

In the next sections we describe the coupling of Dirac and KG field to torsion and discuss entanglement in both cases.

\section{Dirac equation in presence of torsion}\label{sez3}

The Dirac Lagrangian in a RC spacetime can be written as
\be \label{11}
\mathcal{L}_D= - \frac{1}{2} \left[ \bar{\psi} \tilde{\gamma}^\mu D_{\mu} \psi- (D_{\mu} \bar{\psi})
\tilde{\gamma}^\mu \psi \right]- m \bar{\psi} \psi,
\ee
where the covariant derivatives of spinors $\psi$ and their complex conjugates $\bar{\psi}$ are defined as \cite{26,27}
\begin{align}
&D_{\mu} \psi= \tilde{D}_{\mu} \psi- \frac{1}{4}K_{\alpha \beta \mu}\tilde{\gamma}^\alpha \tilde{\gamma}^\beta \psi\,, \label{12}\\
& D_{\mu} \bar{\psi}= \tilde{D}_{\mu} \bar{\psi}+ \frac{1}{4}K_{\alpha \beta \mu} \bar{\psi} \tilde{\gamma}^\alpha \tilde{\gamma}^\beta\,. \label{13}
\end{align}
Here $K_{\alpha \beta \mu}$ is the contorsion tensor in the fully covariant form and $\tilde{D}_{\mu}$ is the covariant derivative of a spinor in a torsionless spacetime. Choosing the FRW metric from \eqref{8} and introducing the tetrad field\footnote{A tetrad is needed when dealing with spinors in a curved spacetime. See, for example, \cite{30}.}
\be \label{14}
e^{\mu}_i= \frac{1}{a(\tau)} \delta^{\mu}_i\,,
\ee
we have that \cite{31}
\be
\tilde{D}_{\mu}\psi= \left(\partial_{\mu}+ \frac{1}{4}\frac{\dot{a}}{a}[\gamma_{\mu}, \gamma^0] \right)\psi,
\ee \label{15}
where the dependence of the scale factor $a$ on $\tau$ is understood from now on.
Here $\gamma^{\mu}$ are the flat gamma matrices, chosen according to the notation of \cite{16}. The curved gamma matrices are defined as
$\tilde{\gamma}^\mu:= e^{\mu}_i\gamma^{i}= a^{-1} \gamma^i$.
Moreover, in a FRW spacetime the torsion tensor assumes the form \eqref{9}.

Accordingly, from the Lagrangian \eqref{11} we obtain the Dirac equation in a FRW spacetime with torsion \cite{26}
\be \label{16}
\left[\frac{\gamma^{\mu}}{a} \left(\partial_{\mu}+ \frac{1}{4} \frac{\dot{a}}{a}[\gamma_{\mu}, \gamma^0] \right)+m  \right] \psi=- \frac{3i}{2} f(\tau)a \gamma^0 \gamma^5 \psi.                        \ee
Using the ansatz \cite{19}
\be \label{17}
\psi=a^{-3/2}(\gamma^{\nu}\partial_{\nu}-M) \varphi
\ee
with $M=ma$, we obtain
\be \label{18}
(\eta^{\mu \nu}\partial_{\mu} \partial_{\nu}- \gamma^0 \dot{M}-M^2)\varphi= - \frac{3i}{2} F(\tau) \gamma^0 \gamma^5 (\gamma^{\nu}\partial_{\nu}-M) \varphi,
\ee
where $F(\tau)=f(\tau)a^2$ and $\eta^{\mu \nu}$ is the Minkowski metric tensor. Eq. \eqref{18} can be solved, in principle, upon specification of the torsion function $f$ and the scale factor $a$.

We assume now an asympotically flat spacetime, with a scale factor of the form
\be \label{19}
a(\tau)=A+B \tanh(\rho \tau),
\ee
widely used for the properties of controlling both the volume and the expansion of the universe \cite{16}. In fact, here $A$ and $B$ are parameters controlling the volume of the universe and $\rho$ the rapidity of expansion.
In the asymptotic (\textit{in} and \textit{out}) regions, Eq. \eqref{18} can be solved with the ansatz
\be \label{20}
\varphi_{\text{in}/\text{out}}= N_{\text{in}/\text{out}} e^{-iE_{\text{in}/\text{out}}\tau} e^{i {\bf p}\cdot {\bf x}} \begin{pmatrix} u_d \\ v_d \end{pmatrix},
\ee
where $N_{\text{in}/\text{out}}$ is a normalization factor and $u_d$, $v_d$ ($d=\uparrow$,$\downarrow$) are two-component spinors, so that
\be \label{21}
\gamma^0 u_d=-iu_d,\ \ \ \ \ \ \gamma^0v_d=iv_d.
\ee
Inserting \eqref{20} into \eqref{18} we obtain the equation
\begin{align} \label{22}
&(E_{\text{in}/\text{out}}^2- \lvert {\bf p} \rvert^2-M_{\text{in}/\text{out}}^2) \begin{pmatrix} u_d \\ v_d \end{pmatrix}= \notag \\
&\frac{3}{2} F_{\text{in}/\text{out}} \begin{pmatrix} \boldsymbol \sigma \cdot {\bf p} & -E_{\text{in}/\text{out}}+M_{\text{in}/\text{out}} \\ -E_{\text{in}/\text{out}}-M_{\text{in}/\text{out}} &  \boldsymbol \sigma \cdot {\bf p} \end{pmatrix} \begin{pmatrix} u_d \\ v_d\end{pmatrix},
\end{align}
where $\boldsymbol \sigma= (\sigma_1, \sigma_2, \sigma_3)$ is the set of Pauli matrices and
\begin{align}
&M_{\text{in}/\text{out}}=m a(\tau \rightarrow -/+ \infty)\,, \label{23} \\
&F_{\text{in}/\text{out}}= f(\tau \rightarrow -/+ \infty) a^2(\tau \rightarrow -/+ \infty)\,.
\end{align}
From Eq. \eqref{22} we can derive the spinor solution with positive energy, that is found to be
\be \label{25}
w_d= \begin{pmatrix} u_d \\ \frac{[(E^2_{\text{in}/\text{out}}- \lvert {\bf p} \rvert^2- M_{\text{in}/\text{out}}^2)- \frac{3}{2} F_{\text{in}/\text{out}}(\boldsymbol \sigma \cdot {\bf p})]}{\frac{3}{2}F_{\text{in}/\text{out}}(-E_{\text{in}/\text{out}}+M_{\text{in}/\text{out}})} u_d \end{pmatrix}
\ee
and similarly for the solution with negative energy, with the substitution $p^{\mu} \rightarrow -p^{\mu}$.
Accordingly, the complete positive-energy solution of the Dirac equation \eqref{18} in the asymptotic regions can be written as
\be \label{26}
U_{\text{in}/\text{out}}({\bf x}, {\bf p}, d, \tau)=N_{\text{in}/\text{out}} (\gamma^{\nu}\partial_{\nu}-M)e^{-iE_{\text{in}/\text{out}}\tau} e^{i {\bf p}\cdot {\bf x}} w_d.
\ee
Imposing the normalization as in \cite{18}, namely $
\bar{U}U= i U^\dagger \gamma^0 U= \delta_{d,d^\prime}$,
one finds
\be \label{28}
N= \frac{\frac{3}{2}F(E-M)}{(E^2-M-\lvert {\bf p} \rvert^2)\sqrt{\left(\frac{3}{2}F\right)^2+ 3F\lvert {\bf p} \rvert +\lvert {\bf p} \rvert^2-(E-M)^2}},
\ee
where the subscripts in/out have been omitted for brevity.

The only missing element is now the energy correction due to the presence of torsion. We write the total energy as
\be \label{29}
E_{\text{in}/\text{out}}=E_{0}+x=\sqrt{ \lvert {\bf p} \rvert^2+M_{\text{in}/\text{out}}}+x
\ee
where $E_{0}$ is the energy when torsion is not present and $x$ is the correction due to the torsion contribution. Inserting \eqref{29} into \eqref{22} and computing the determinant of the corresponding matrix, one finds
\be \label{30}
E^{\pm}= \sqrt{\lvert {\bf p} \rvert^2+ M^2+ \frac{\left(\frac{3}{2}F\right)^2 \pm \sqrt{\left(\frac{3}{2}F\right)^4+8\lvert {\bf p}^2 \rvert \left(\frac{3}{2}F\right)^2}}{2}},
\ee
where again we have omitted the subscript in/out to simplify the notation. Since we expect that energy corrections due to torsion are typically small, the expression \eqref{30} can be simplified to
\be \label{31}
E^{\pm}=E_{0}+ \frac{\left(\frac{3}{2}F \right)^2\pm \sqrt{\left(\frac{3}{2}F\right)^4+8\lvert {\bf p}^2 \rvert \left(\frac{3}{2}F\right)^2}}{4 E_{0}}.
\ee
Clearly, for anti-particles the ansatz would be $E_{\text{in/out}}=-E_{0}+x$ and so one finds the opposite of Eq. \eqref{31}. Moreover, if we assume a positive torsion field, the solution $E^+$ should be excluded, in order to assure that \eqref{25} is a positive-energy spinor. Analogously, if we assume  a negative torsion field, we should exclude $E^-$ for the same reason.

\subsection{Particle creation and entanglement}

To study entanglement for Dirac field in a FRW spacetime with torsion, we should be able to compute the Bogolyubov coefficients that relates the in and out regions \cite{16,18,19}. However, this cannot be done analytically, since the Dirac equation \eqref{16} can be solved only in the two asymptotic regions separately.

In order to  qualitatively understand how entanglement is affected by torsion, we can imagine that the torsion field $f(\tau)$ is not involved in the dynamics of Dirac field, i.e., it becomes negligible during the expansion of the universe.
This assumption can be justified if we recall that torsion is typically relevant only when high mass densities are present \cite{32}, as happened in the early universe. Accordingly, a suitable form for torsion function might be
\be \label{32}
F(\tau)=  F_0\,a^{-n},\ \ \ \ \ n \in \mathbb{N}
\ee
and so
\be \label{33}
f(\tau)= f_0\,a^{-k},\ \ \ \ \ k \in \mathbb{N},\  k \geq 3.
\ee
The constant $f_0$ should assume values much smaller than the mass $m$ (natural units).
Moreover, we assume charge and angular momentum conservation, as in \cite{19}. With this assumption, the Bogolyubov transformations that relate the \textit{in} and \textit{out} creation and destruction operators can be written as \cite{19}
\be \label{34}
\begin{aligned}
&a_{\text{in}}({\bf p},d)= \mathcal{A}^*(p)a_{\text{out}}({\bf p},d)+ \beta_{d,-d}^*({\bf p}) b_{\text{out}}^\dagger({\bf p},-d) \\
&b_{\text{in}}^\dagger(-{\bf p},d)=- \beta_{-d,d}({\bf p})a_{\text{out}}({\bf p},-d)+\mathcal{A}(p) b_{\text{out}}^\dagger(-{\bf p},d)
\end{aligned}
\ee
where $p\equiv \lvert p \rvert$. Here $a_{\text{in}}, b_{\text{in}}$ and $a_{\text{out}}, b_{\text{out}}$ are the annihilation operators of particles and anti-particles in the \textit{in} and \textit{out} regions, respectively. The coefficient $\mathcal{A}(p)$ becomes \cite{16}
\be \label{35}
\mathcal{A}(p)= \sqrt{\frac{M_{\text{in}}}{M_{\text{out}}} \frac{E_{\text{in}}}{E_{\text{out}}}} \frac{N_{\text{in}}}{N_{\text{out}}} A(p)
\ee
and it can be considered real, without loss of generality \cite{19}. Moreover, from the algebra of fermionic operators it turns out that $\lvert \mathcal{A}(p) \rvert^2+ \lvert \beta_{d,-d}({\bf p}) \rvert^2=1$.

If the torsion term is negligible during the expansion, the coefficient $A(p)$ can be determined resorting to Hypergeometric functions \cite{15,16}. One thus gets
\be \label{36}
A(p)= \frac{\Gamma(1-(i/\rho)E_{\text{in}})\Gamma(-(i/\rho) E_{\text{out}})}{\Gamma(1-(i/\rho) E_{+} - imB/\rho)\Gamma(-(i/\rho) E_{+}+imB/\rho) },
\ee
where $\Gamma(x)$ is the usual gamma function and
\be \label{37}
E_{\pm}\equiv \frac{1}{2}(E_{\text{out}} \pm E_{\text{in}}).
\ee
Inverting Eq. \eqref{34}, we can compute the number $n$ of particles per mode, created due to the universe expansion \cite{19}
\begin{align}
&n^p(p,\uparrow)=\langle 0_{\text{in}} \rvert a^\dagger_{\text{out}}({\bf p}, \uparrow) a_{\text{out}}({\bf p}, \uparrow) \lvert 0_{\text{in}} \rangle= \lvert \beta_{\downarrow \uparrow} \rvert^2, \label{38}\\
&n^p(p,\downarrow)= \langle 0_{\text{in}} \rvert a^\dagger_{\text{out}}({\bf p}, \downarrow) a_{\text{out}}({\bf p}, \downarrow) \lvert 0_{\text{in}} \rangle= \lvert \beta_{\uparrow \downarrow} \rvert^2, \label{39}
\end{align}
and analogously for anti-particles. The unitary operator acting on the Fock space and representing the transformation \eqref{34} has been derived in \cite{19} and so, applying it to the \textit{out} vacuum state, we get
\begin{widetext}
\begin{align} \label{40}
\lvert 0_p; 0_{-p} \rangle_{\text{in}}= \mathcal{A}^2 \bigg( \lvert 0_p; 0_{-p} \rangle_{\text{out}}- \frac{\beta_{\uparrow \downarrow}^*}{\mathcal{A}} \lvert \uparrow_p ; \downarrow_{-p} \rangle_{\text{out}} - \frac{\beta_{\downarrow \uparrow}^*}{\mathcal{A}} \lvert \downarrow_p ; \uparrow_{-p} \rangle_{\text{out}} +\frac{\beta^*_{\uparrow \downarrow} \beta^*_{\downarrow \uparrow}}{\mathcal{A}^2} \lvert \uparrow \downarrow_p; \uparrow \downarrow_{-p} \rangle_{\text{out}} \bigg).
\end{align}
\end{widetext}

The particle-antiparticle density operator corresponding to Eq.  \eqref{40} in the \textit{out} region will be
\be \label{41}
\rho^{(\text{out})}_{p,-p}= \lvert 0_p; 0_{-p} \rangle_{\text{in}} \langle 0_p; 0_{-p} \rvert,
\ee
and taking the partial trace over anti-particles we obtain the reduced density operator
\begin{align} \label{42}
\rho_p^{(\text{out})}=&\text{Tr}_{-p}(\rho^{(\text{out})}_{p,-p})= \mathcal{A}^4 \lvert 0_p \rangle \langle 0_p \rvert + \mathcal{A}^2 \lvert \beta_{\uparrow \downarrow} \rvert^2 \lvert \uparrow_p \rangle \langle \uparrow_p \rvert \notag \\[5 pt]
&+ \mathcal{A}^2 \lvert \beta_{\downarrow \uparrow} \rvert^2 \lvert \downarrow_p \rangle \langle \downarrow_p \rvert+ \lvert \beta_{\uparrow \downarrow} \rvert^2 \lvert \beta_{\downarrow \uparrow} \rvert^2 \lvert \uparrow \downarrow_p \rangle \langle \uparrow \downarrow_p \rvert.
\end{align}
If we assume now that
\be \label{43}
n^p(p,\uparrow)=n^p(p,\downarrow)=n^a(p,\uparrow)= n^a(p,\downarrow)= \frac{n(p)}{4},
\ee
we obtain that the coefficients in Eq. \eqref{42} solely depend on $n$, that is
\be \label{44}
\mathcal{A}^2= \frac{4-n(p)}{4},\ \ \ \ \lvert \beta_{\uparrow \downarrow} \rvert^2= \lvert \beta_{\downarrow \uparrow}   \rvert^2= \frac{n(p)}{4}.
\ee
To evaluate the amount of particle-antiparticle entanglement of Eq. \eqref{42}, we can use the subsystem entropy \cite{19}, since the state \eqref{41} is pure. Accordingly, we can write
\begin{align} \label{45}
S\left( \rho_p^{(\text{out})} \right)= &-2 \left(\frac{4-n}{4} \right) \log_2\left(\frac{4-n}{4} \right) \notag \\
&-2 \left( \frac{n}{4} \right) \log_2 \left( \frac{n}{4} \right),
\end{align}
where the dependence of $n$ on the momentum $p$ is understood.

\begin{figure}[ht!]
    \centering
    \includegraphics[scale=0.32]{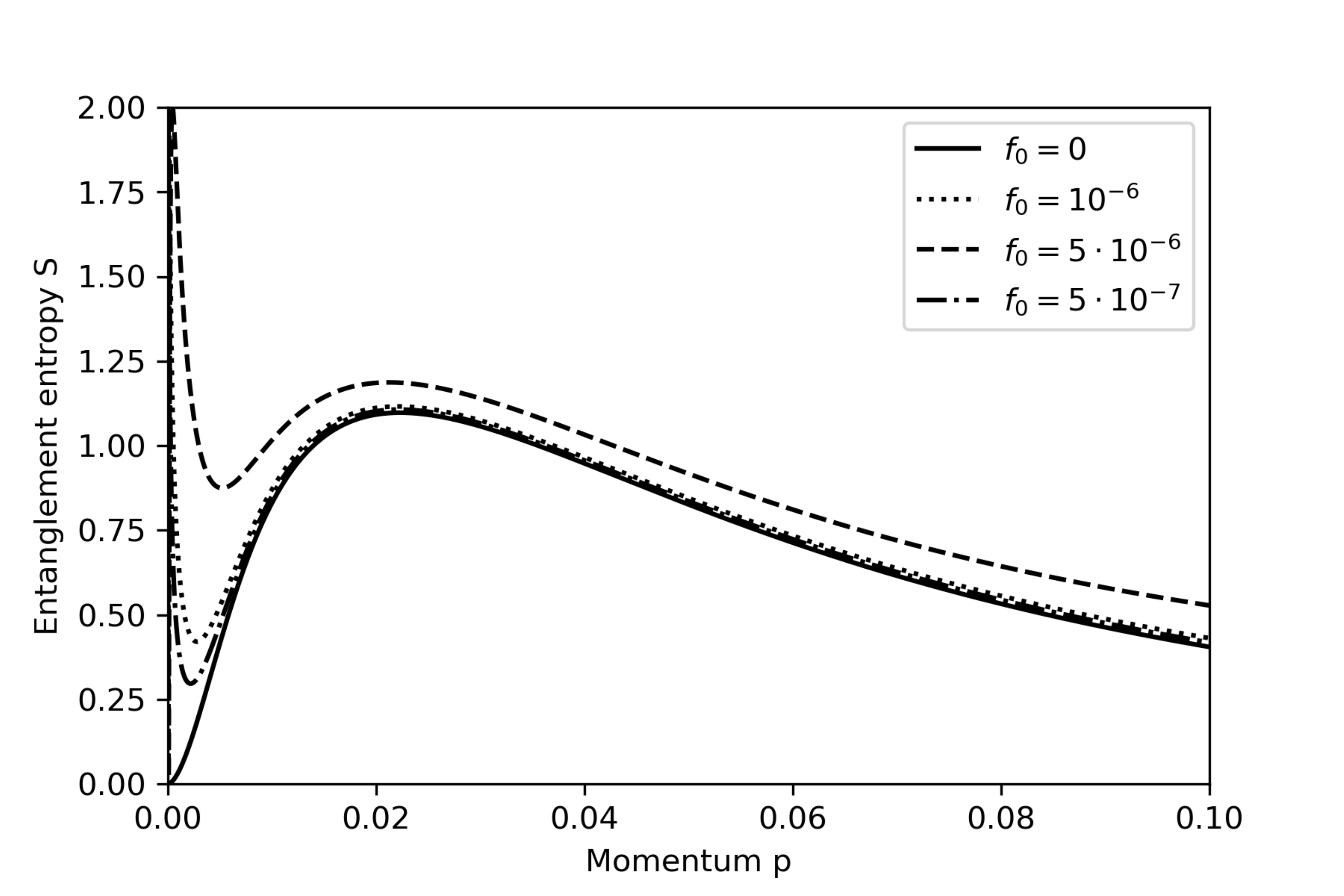}
    \caption{Entanglement entropy for Dirac field in presence of torsion. The values of the parameters are: $m=0.01$, $A=3$, $B=2$, $\rho=1$ and $k=6$.}
    \label{fig1}
\end{figure}
Plotting the entropy $S$ as $p$ varies, we see that torsion is expected to increase the amount of entanglement for Dirac field. This is true in particular for small values of the momentum $p$.

\begin{figure}
    \centering
    \includegraphics[scale=0.35]{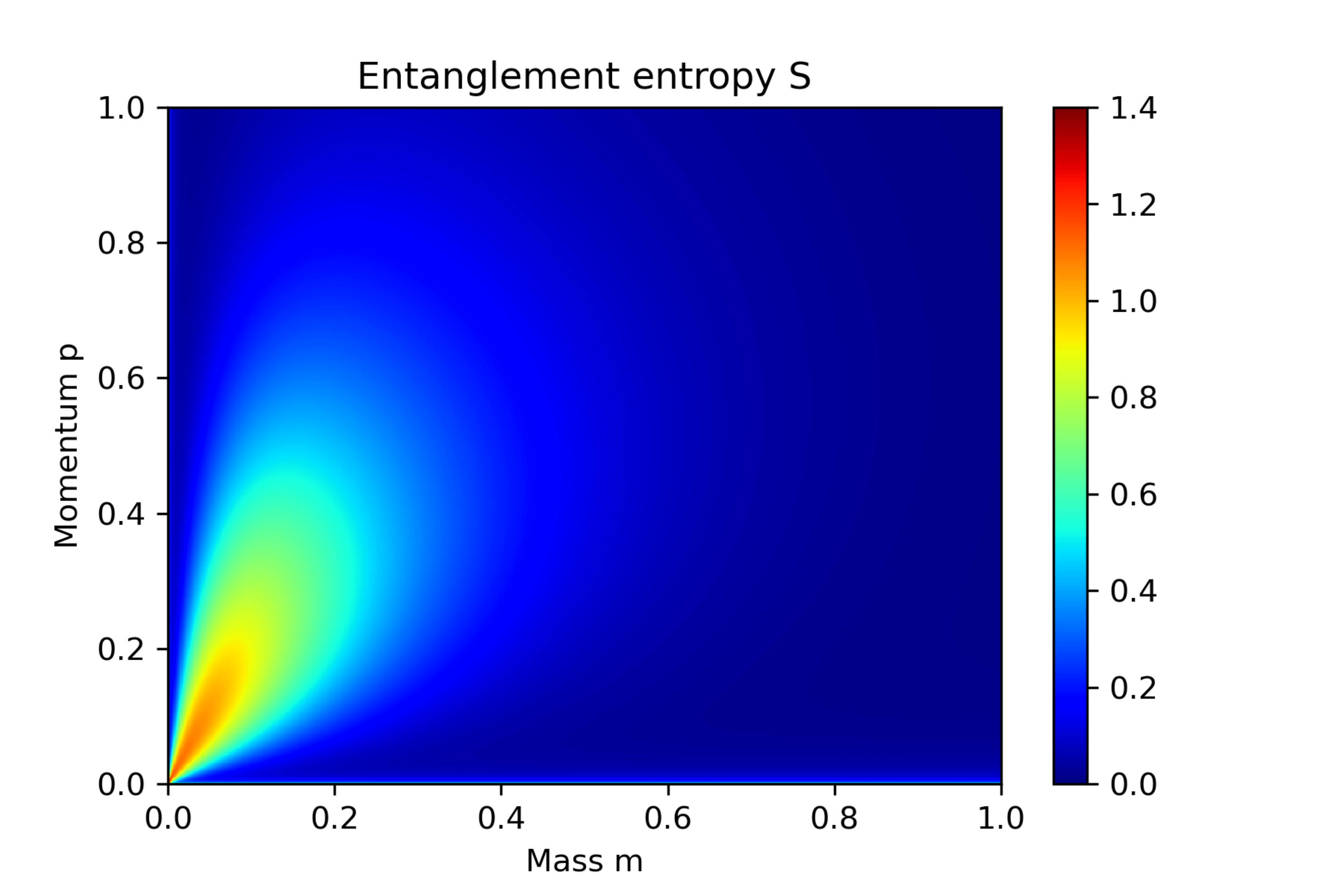}
    \caption{Entanglement entropy when both the field parameters m and $\lvert {\bf p} \rvert=p$ are varied. The other parameters are: $f_0=10^{-6}$, $A=2$, $B=3$, $\rho=1$ and $k=6$.}
    \label{fig2}
\end{figure}

When the mass $m$ increases, the corrections to $S$ due to torsion are almost indistinguishable, as can be seen from Fig. \ref{fig2}. This happens because in this case the energy corrections due to torsion becomes even smaller with respect to the energy without torsion $E_{0}$.

\section{KG equation in presence of torsion}\label{sez4}

In a curved spacetime, described by the metric $g_{\mu \nu}$, the Lagrangian leading to KG equation can be written as \cite{16}
\be \label{46}
\mathcal{L}_{KG}= - \frac{1}{2}  g^{\mu \nu} \partial_{\mu} \phi\  \partial_{\nu}\phi- \frac{1}{2}  (m^2+ \xi R)\phi^2.
\ee
Here $m$ is the mass of the bosonic field $\phi$, while the term $\xi R \phi^2$ describes the coupling of the field to the curvature of spacetime. Two values of $\xi$ are of particular interest: $\xi=0$ (minimal coupling) and $\xi=1/6$ (conformal coupling).

The Lagrangian \eqref{46} leads to the KG equation
\be \label{47}
\frac{1}{\sqrt{-g}} \partial_{\mu}\big(\sqrt{-g}\  g^{\mu \nu} \partial_{\nu} \phi \big)- \big( m^2+ \xi R \big) \phi=0.
\ee
As expected, torsion cannot be minimally coupled to KG field, since the covariant derivatives of general relativity reduces to partial derivatives in the scalar case. However, torsion appears implicitly in the Ricci-Cartan scalar curvature $R$.

Specializing now to the case of our conformally flat metric \eqref{8}, we obtain
\be \label{48}
\frac{1}{a^2} \square \phi - \frac{2 \dot{a}}{a^3} \dot{\phi}- \big( m^2+ \xi R \big) \phi=0,
\ee
where $\square$ is the usual D'Alembertian operator. If torsion is present, in the form described by Eq. \eqref{9}, we can use Eq.  \eqref{3} to obtain the scalar curvature in FRW spacetime with torsion,
\be \label{49}
R= 6 \bigg[\frac{ \ddot{a}}{a^3}- \frac{ f^2(\tau)}{a^6}- \frac{2\dot{h}(\tau)}{a^2} - \frac{ 4\dot{a}}{a^3} h(\tau)+ \frac{4 h^2(\tau)}{a^2} \bigg].
\ee
Assuming now that conformal coupling holds ($\xi=1/6$), Eq. \eqref{48} becomes
\begin{align} \label{50}
\frac{1}{a^2} \square \phi - \frac{2 \dot{a}}{a^3} \dot{\phi}- \bigg[ m^2 + \frac{\ddot{a}}{a^3}- \frac{f^2(\tau)}{a^6} - \frac{2\dot{h}(\tau)}{a^2} \notag \\
- \frac{ 4\dot{a}}{a^3} h(\tau)+ \frac{4 h^2(\tau)}{a^2} \bigg] \phi=0.
\end{align}
 This equation may be further simplified by making the substitution $\phi \rightarrow \chi= a \phi$, to give
\be \label{51}
\square \chi =\bigg[ a^2m^2 - \frac{f^2(\tau)}{a^4} - 2\dot{h}(\tau) -  \frac{4\dot{a}}{a} h(\tau)+4 h^2(\tau) \bigg] \chi.
\ee
The general solution of Eq. \eqref{51} can be written in the form \cite{16}
\be \label{52}
\chi_{p}({\bf x}, \tau)= e^{i {\bf p} \cdot {\bf x}}\chi_p(\tau),
\ee
where $\chi_{\bf p}(\tau)$ satisfies the following differential equation
\begin{widetext}
\begin{align} \label{53}
\ddot{\chi}_p(\tau)+ \bigg[&\lvert {\bf p} \rvert^2+m^2a^2 - \frac{f^2(\tau)}{a^4} - 2\dot{h}(\tau)-  \frac{4\dot{a}}{a} h(\tau)+4 h^2(\tau)\bigg]\chi_p(\tau)=0.
\end{align}
\end{widetext}

This equation can be solved exactly in some particular cases. If we assume that the scale factor is \eqref{19}, as in the Dirac case, a solution can be found if we assume
\begin{align} \label{54}
&f(\tau)=f_0a^3(\tau)
&h(\tau)=h_0a(\tau),
\end{align}
with $f_0$, $h_0$ constants. In particular, we are interested in the asymptotic solutions $\chi^{\text{in}}_p$, $\chi^{\text{out}}_p$, that can be written as
\begin{widetext}
\begin{align}
&\chi_p^{\text{in}}(\tau)=\exp\left\{ -i \left[\mathcal{E}_+\tau+ \frac{1}{\rho}\mathcal{E}_- \ln(2 \cosh(\rho \tau) \right] \right \}
\tensor*[^{}_2]{F}{^{}_1}\bigg(1+ \frac{i}{\rho}\mathcal{E}_{-}-\frac{6h_0B}{\rho}, \frac{i}{\rho}\mathcal{E}_{-}+ \frac{6h_0B}{\rho};1- \frac{i}{\rho}\mathcal{E}_{\text{in}}; \frac{1+ \tanh(\rho\tau)}{2} \bigg)\,,\label{55}\\
\,\nonumber\\
&\chi_p^{\text{out}}(\tau)=\exp\left\{ -i \left[\mathcal{E}_+\tau+ \frac{1}{\rho}\mathcal{E}_- \ln(2 \cosh(\rho \tau) \right] \right \}  \tensor*[^{}_2]{F}{^{}_1}\bigg(1+ \frac{i}{\rho}\mathcal{E}_{-}-\frac{6h_0B}{\rho}, \frac{i}{\rho}\mathcal{E}_{-}+ \frac{6h_0B}{\rho}; 1+ \frac{i}{\rho}\mathcal{E}_{\text{out}}; \frac{1- \tanh(\rho\tau)}{2} \bigg)\,,\label{56}
\end{align}
\end{widetext}
where $\tensor*[^{}_2]{F}{^{}_1}$ is the Hypergeometric function of second kind and we have introduced
\begin{align}
&\mathcal{E}_{\text{in}}= \big[ \lvert {\bf p} \rvert^2 +(m^2-f_0^2+4h_0^2)a^2(\tau \rightarrow -\infty) \big]^{1/2}, \\
&\mathcal{E}_{\text{out}}= \big[ \lvert {\bf p} \rvert^2 +(m^2-f_0^2+4h_0^2)a^2(\tau \rightarrow +\infty) \big]^{1/2}.
\end{align}
The quantities $\mathcal{E}_{\pm}$ are defined as in Eq. \eqref{37}.

\subsection{Particle creation and entanglement}
It has already been shown that a dynamical spacetime generates entanglement between particle ($p$) and antiparticle ($-p$) modes of a KG field \cite{33}. Here we revisit the mechanism that leads to entanglement, assuming the presence of torsion.

Following the standard quantization procedure, we associate to each mode $\chi_p^{\text{in/out}}({\bf x}, \tau)$ and to its complex conjugate $\chi_p^{\text{in/out}*}({\bf x}, \tau)$ annihilation and creation operators $a_{\text{in/out}}({\bf p})$, $a_{\text{in/out}}^\dagger({\bf p})$. These operators satisfies equal-time commutation relations \cite{33,34} and the two sets of modes define two representations of the scalar field \cite{16}
\begin{align} \label{59}
\chi({\bf x}, \tau)&= \int \frac{d^3p}{(2\pi)^3} \frac{1}{[2 \mathcal{E}_{\text{in}}]^{1/2}} \notag \\
&\ \ \ \ \ \ \ \  [\chi_p^{\text{in}}({\bf x}, \tau)a_{\text{in}}({\bf p})+\chi_p^{\text{in}*}({\bf x}, \tau)a_{\text{in}}^\dagger({\bf p})]
\notag \\[6 pt]
&= \int \frac{d^3p}{(2\pi)^3} \frac{1}{[2 \mathcal{E}_{\text{out}}]^{1/2}} \notag \\
&\ \ \ \ \ \ \ \  [\chi_p^{\text{out}}({\bf x}, \tau)a_{\text{out}}({\bf p})+\chi_p^{\text{out}*}({\bf x}, \tau)a_{\text{out}}^\dagger({\bf p})].
\end{align}
Expanding now one mode in terms of the other
\be \label{60}
\chi_k^{\text{in}}({\bf x}, \tau)= \alpha(p) \chi_p^{\text{out}}({\bf x}, \tau)+ \beta(p) \chi_{-p}^{\text{out}*}({\bf x}, \tau)\,,
\ee
and inserting this expression into Eq. \eqref{59}, we obtain a map between \text{in} and \text{out} operators:
\be \label{61}
a_{\text{out}}({\bf p})= \left(\frac{\mathcal{E}_{\text{out}}}{\mathcal{E}_{\text{in}}}\right)^{1/2}[\alpha(p) a_{\text{in}}({\bf p})+ \beta^*(p) a_{\text{in}}^\dagger(-{\bf p})].
\ee
The coefficients $\alpha(p),  \beta(p)$ are the Bogolyubov coefficients for this transformation. From the commutation relations for bosonic operators, we have \cite{16}
\be \label{62}
\lvert \alpha(p) \rvert^2- \lvert \beta(p) \rvert^2= \frac{\mathcal{E}_{\text{in}}}{\mathcal{E}_{\text{out}}}.
\ee
Recalling the asymptotic solutions \eqref{55} and \eqref{56}, the Bogolyubov coefficients $\alpha(p)$ and $\beta(p)$ follow from the linear transformation properties of Hypergeometric functions \cite{16}. We have
\begin{align} \label{63}
&\alpha(p)= \dfrac{\Gamma(1-(i/\rho)\mathcal{E}_{\text{in}}) \Gamma(-(i/\rho)\mathcal{E}_{\text{out}})}{\Gamma(1-(i/\rho)\mathcal{E}_+-6hB/\rho)\Gamma(-(i/\rho)\mathcal{E}_{+}+6hB/\rho)}\,,\nonumber\\
\,\\
&\beta(p)=  \dfrac{\Gamma(1-(i/\rho)\mathcal{E}_{\text{in}}) \Gamma((i/\rho)\mathcal{E}_{\text{out}})}{\Gamma(1+(i/\rho)\mathcal{E}_{-}+6hB/\rho)\Gamma((i/\rho)\mathcal{E}_{-}-6hB/\rho)}\,.\nonumber
\end{align}
Now, let us suppose that the KG field is in the vacuum state of the \textit{in} modes, $\lvert 0 \rangle_{\text{in}}$, and we want to evaluate the expectation value of the particle number operator for the \textit{out} modes. We simply have to insert Eq. \eqref{61} and its complex conjugate into the expression $_{\text{in}}\langle 0 \rvert a_{\text{out}}^\dagger({\bf p}) a_{\text{out}}({\bf p}) \rvert 0 \rangle_{\text{in}}$, finding
\be \label{64}
_{\text{in}}\langle 0 \rvert a_{\text{out}}^\dagger({\bf p}) a_{\text{out}}({\bf p}) \rvert 0 \rangle_{\text{in}}= \lvert \beta(p) \rvert^2.
\ee
Thus, the vacuum \textit{in} state is not empty in the \textit{out} region and $\lvert \beta(p) \rvert^2$ is interpreted as the number of detected quanta in the mode $p$.

\begin{figure}[ht!]
    \centering
    \includegraphics[scale=0.28]{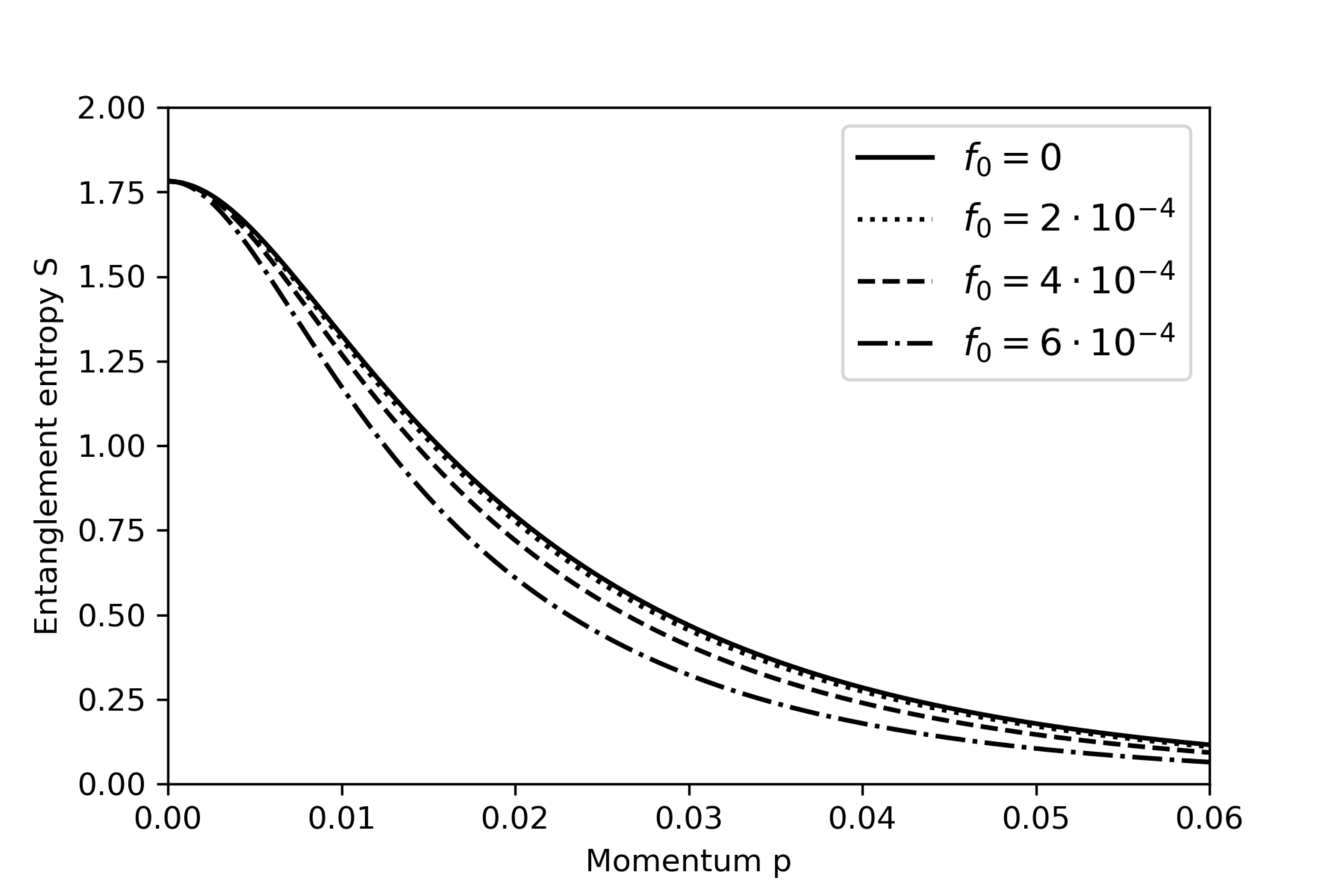}
    \caption{Entanglement entropy for KG particles for different values of the torsion parameter $f_0$. The other parameters are: $m=0.01$, $h_0=10^{-6}$, $A=3$, $B=2$ and $\rho=1$.}
    \label{fig3}
\end{figure}

\begin{figure}[ht!]
    \centering
    \includegraphics[scale=0.28]{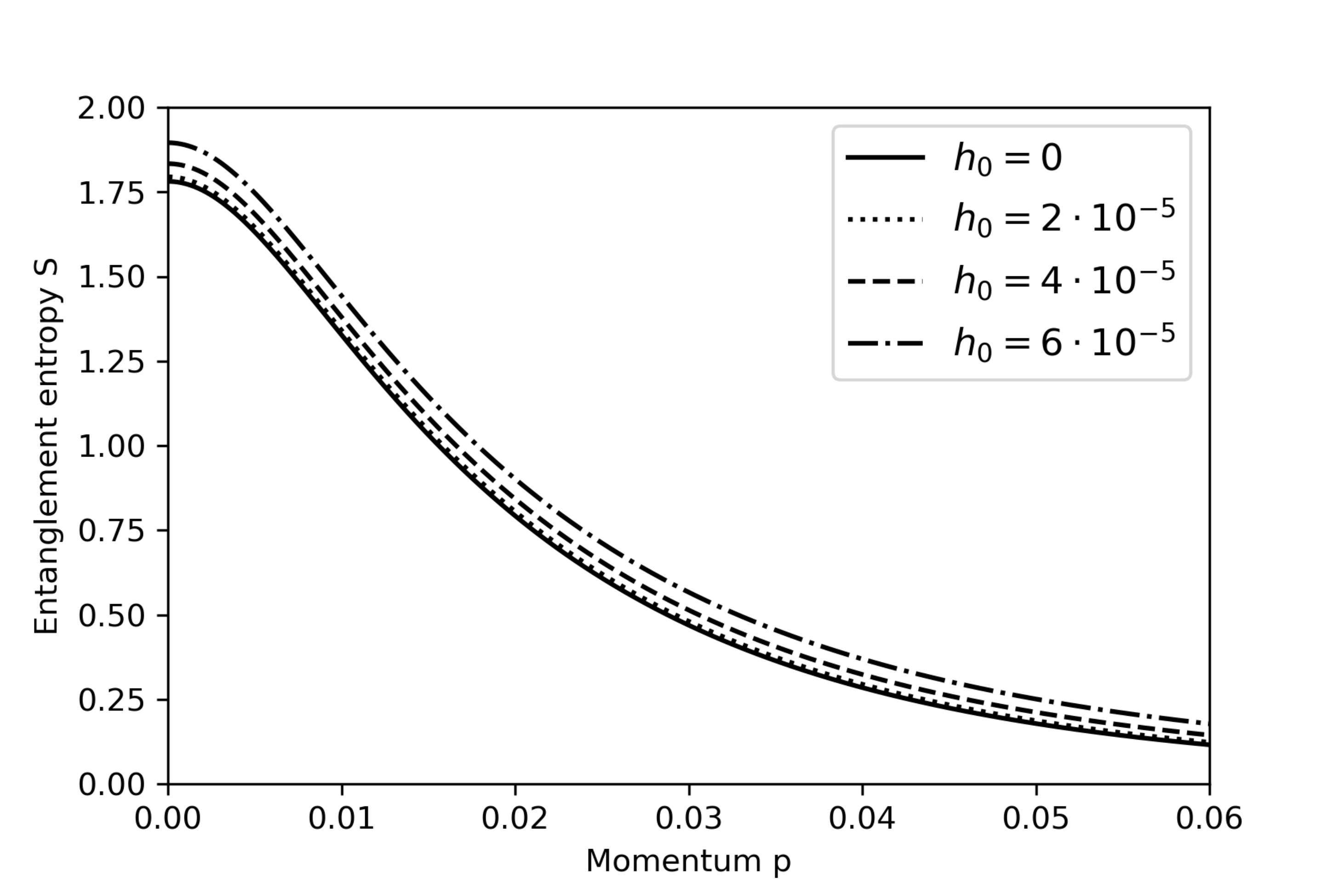}
    \caption{Entanglement entropy for KG particles for different values of the torsion parameter $h_0$. The other parameters are: $m=0.01$, $f_0=10^{-6}$, $A=3$, $B=2$ and $\rho=1$.}
    \label{fig4}
\end{figure}

To discuss entanglement, we write the \textit{in} vacuum as a Schmidt decomposition of \textit{out} states
\be \label{65}
\lvert 0_p ; 0_{-p} \rangle_{\text{in}}= \sum_{n=0}^{\infty} c_n \lvert n_p;n_{-p} \rangle_{\text{out}},
\ee
where the Schmidt coefficients are  \cite{33}
\be \label{66}
c_n=\left(\frac{\beta^*(p)}{\alpha^*(p)}\right)^nc_0,
\ee
with
\be \label{67}
c_0=\sqrt{1- \left \lvert \frac{\beta^*(p)}{\alpha^*(p)} \right \rvert^2}.
\ee
From the state \eqref{65} we can write the bipartite density matrix
\be \label{68}
\rho^{\text{(out)}}_{p,-p}= \lvert 0_p ; 0_{-p} \rangle_{\text{in}} \langle 0_p ; 0_{-p}  \rvert.
\ee
Since the Schmidt coefficients \eqref{66} are non-zero, it follows that the \textit{in} vacuum is entangled from the point of view of an \textit{out} observer. As for the Dirac case, the amount of particle-antiparticle entanglement is quantified considering the reduced density matrix
\be \label{69}
\rho^{\text{(out)}}_{p}= \text{Tr}_{-p}(\rho^{\text{(out)}}_{p,-p})= \sum_{m=0}^{\infty}\  _{\text{out}}\langle m_{-p} \rvert \rho^{\text{(out)}}_{p,-p} \rvert m_{-p} \rangle_{\text{out}}.
\ee
Accordingly, the Von Neumann entropy of this state takes the form
\begin{align} \label{70}
S(\rho^{\text{(out)}}_{p})&= - \text{Tr}\left(\rho^{\text{(out)}}_{p} \log_2 \rho^{\text{(out)}}_{p}  \right) \notag \\
&= \log_2 \frac{\gamma^{\gamma/(\gamma-1)}}{1-\gamma},
\end{align}
where \cite{33}
\be \label{71}
\gamma= \left \lvert \frac{\beta(p)}{\alpha(p)} \right \rvert^2= \frac{\sinh^2(\pi \mathcal{E}_{-}/\rho)}{\sinh^2(\pi \mathcal{E}_+/\rho)}.
\ee
In Figs. \ref{fig3} and \ref{fig4} we show how KG entanglement is affected by the presence of the parameters $f_0$ and $h_0$. In particular, if $h_0$ is non-zero, the amount of entanglement is increased, while a non-zero $f_0$ modifies the mode dependence of $S$.

\section{Discussion on physical consequences related to torsion}\label{sez5}

In the case of Dirac field, the entanglement entropy $S$ is upper bounded and no divergences occur. More precisely, $S$ is bounded by $S < \log_2N$, where $N$ is the Hilbert space dimension of the partial state (reduced density operator). In our picture, the hypothesis of charge and angular momentum conservation \cite{33,34} still holds, so from Eq. \eqref{42} we see that $N=4$. Accordingly, the maximum  value for the entropy is $S=2$. In presence of torsion, we see that the entropy provides two maxima: the first is absolute and corresponds to $p=0$, whereas the second is relative and lies around $p\simeq 2m$. The two maxima are portrayed in Figs. \ref{fig1} and \ref{fig2}. The relative maximum can be interpreted in view of the Pauli exclusion principle. Indeed, it is not possible to \emph{condensate} fermions as $p\rightarrow0$ and consequently there would exist a $p\neq0$ at which the maximum occurs. The absolute maximum could be interpreted in view of the torsion field, adopted in Eq. \eqref{33}. The function $f(\tau)$ is constructed to guarantee the cosmological principle to hold. Even though appealing, this choice disagrees with the case of particles with half-integer spin that, by virtue of the Pauli exclusion principle,  cannot occupy the same quantum state within a quantum system simultaneously. Thus, at $p\rightarrow0$ the particle contribution becomes negligibly small, albeit the torsion source due to $f_0$ does not, by construction. This implies that at small $p$ the main contribution is mainly due to torsion, interpreting such a result as direct consequence of the underlying torsion field. The case without torsion stresses our interpretation. In fact, we here have no torsion domination at $p\rightarrow0$ and so $S$ goes to zero as expected by construction. We believe this apparent issue can be healed if the torsion field is chosen to fulfill simultaneously the Pauli exclusion principle and the cosmological principle, extending our treatment by means of a refined one. We will face our hypothesis in future works, proposing alternative versions of the torsion field.

In our scenario, since  $m\neq0$, the fact that the relative maximum is around $p\simeq2m$ is in agreement with our previous discussion because the relative maximum occurs in a region that is far from $p\rightarrow0$. In the region far from $p=0$, the Dirac field seems to resemble the KG framework. However, in the former case we have an absolute maximum that always occurs at $p=0$ and the similarity between the two cases, Dirac and KG curves, is only apparent. Indeed, we believe this apparent similitude is a consequence of the employed torsion field whose functional form is simplified to guarantee the cosmological principle holds. Moreover, again this can be interpreted by the fact that for bosons we do not have any Pauli exclusion principle and at $p=0$ bosons can \emph{condensate} in the fundamental state to provide the maximum plotted in Figs. \ref{fig3} - \ref{fig4}.

In the KG field, we first remark  that the entanglement entropy is not necessarily bounded, due to the infinite dimension corresponding to the density operator \eqref{69}. At $p=0$, the effects due to the torsion field are inferred from a non-minimal coupling between the field and torsion itself. Consequently torsion does not dominate in any regions of $p$ space, differently from the Dirac case.

Comparing the cases with and without torsion suggests that the shapes of each curve continue being similar, albeit slightly different. This is direct consequence of the non-minimal coupling above discussed. The most important fact is that for small values of $f_0$, it seems that $S$ weakly decreases. The opposite happens for $h_0$, i.e. for small values of $h_0$ the entropy appears larger than the case without torsion. These two evidences can be interpreted in view of Eqs. \eqref{54}. Indeed, by construction $f(\tau)$ and $h(\tau)$ scale as the volume and radius of the universe, respectively. Consequently, from Eq. \eqref{49} the term $\propto {f^2\over a^6}$ is a constant throughout the universe evolution, indicating that the curvature is weakly influenced by $f_0$. This implies that the entropy should be smaller for $f_0\neq0$ than the case without torsion. On the other hand, the same does not happen for $h$ since it couples to the term $\propto a^{-2}$ but also to $\propto \dot a$. Moreover, the kinematic term $\dot h$ is also different from zero, involving the fact that as the universe radius  increases, then its contribution increases as well. This acts on the entropy that is larger  than the case without torsion, for $h_0\neq0$.

In all the aforementioned cases, we underline our findings are in line with previous results found in the literature, certifying that the role of torsion modifies the entanglement measure depending on how it couples with the universe expansion history.

\section{Final remarks}\label{sez6}

In this paper, we investigated particle production and entanglement in the framework of EC theory with fermionic and bosonic fields. Thus, we considered the Dirac and KG equations, solving them when particle spin is not negligible. In the framework of FRW universe, we took the most general form for the torsion source, whose constraints are imposed in agreement with the cosmological principle.

We showed how torsion affects entanglement in the cases that enabled us to get analytical solutions in the KG field. Even though we demonstrated no analytical entanglement entropy could be obtained for fermions, assuming torsion to be small enough we got approximate solutions, extending the results when torsion is zero. According to our findings, we showed which properties should be fulfilled by torsion field to get entanglement increase throughout universe's expansion history. In particular, a positive torsion field is required to increase the amount of entanglement for fermions. In this case we also noticed that the mode dependence of the entanglement entropy is drastically modified for small values of the particle momentum.  For the KG field, the amount and mode dependence of entanglement is slightly modified by the two external functions describing torsion. We interpreted the maxima of entanglement for both Dirac and KG fields. In particular, we showed that the Pauli principle is responsible for the relative maximum in the Dirac case, while the absolute maxima for Dirac and KG are direct consequence of the torsion field, involved in our treatment.

Future works will be devoted to understand how one can relate torsion to dark constituents. Moreover, it would be intriguing to compare the various approaches to torsion sources and to apply them to our quantum scenario, checking whether entanglement is modified accordingly. Another interesting avenue of research is related to entanglement extraction from the field modes, using local detector couplings. In particular, it has been shown that modulating a detector's resonance frequency and interaction strength can be useful to optimize the extracted entanglement \cite{43}. In this direction, it would be important to understand what time dependence of interaction would optimize the extraction of information about cosmological parameters and spacetime structure, using this method. This may help to further elucidate the relevance of torsion in the universe history. Finally, future developments are expected from quantum emulation of the universe expansion by means of analogue experiments. In particular, using ion traps it has been shown that ions manifest actual phonon production, if the trap is expanded over a finite time \cite{44}. Moreover, Bose-Einstein condensate models have been proposed \cite{45} to simulate complex inflationary scenarios. There, torsion is expected to play a relevant role and so developing a reliable, robust, and highly tunable laboratory testbed for analogue inflation would be of great experimental value to discuss the role of torsion in cosmology.

\vspace{1cm}

\acknowledgments
O.L. wants to express his gratitude to Kuantay Boshkayev and Manuel Hohmann for fruitful discussions on the main topic of this work. O.L. acknowledges the support of the Ministry of Education and Science of the Republic of Kazakhstan, Grant IRN AP08052311.

\end{document}